\documentclass[12pt, final]{article}
\usepackage[utf8]{inputenc}

\usepackage[round,sort,colon,authoryear]{natbib}
\usepackage{amsmath, amssymb, amsfonts}
\usepackage[colorlinks=true, allcolors=blue]{hyperref}

\usepackage{authblk}

\usepackage{amsmath}
\usepackage{amssymb}
\usepackage{graphicx}
\usepackage{subcaption}

\usepackage{a4wide}

\usepackage[bottom=3cm]{geometry}

\title{Dispersive and non-dispersive nonlinear long wave transformations: Numerical and experimental results}

\author[1]{T. Torsvik}
\author[2]{A. Abdalazeez}
\author[3]{D. Dutykh}
\author[4]{P. Denissenko}
\author[2,5]{I. Didenkulova}

\affil[1]{Norwegian Polar Institute, Fram Centre, Postbox 6606 Langnes, 9296 Troms\o{}, Norway}
\affil[2]{Department of Marine Systems, Tallinn University of Technology, Akadeemia tee 15, Tallinn 12618, Estonia}
\affil[3]{Univ. Grenoble Alpes, Univ. Savoie Mont Blanc, CNRS, LAMA, 73000 Chambéry, France}
\affil[4]{Department of Engineering, University of Warwick, Coventry, CV4 7AL}
\affil[5]{Nizhny Novgorod Technical University n.a. R.E. Alekseev, Minin str. 24, Nizhny Novgorod 603006, Russia}

\date{March 2019}

\begin{document}

\maketitle

\begin{abstract}
    The description of gravity waves propagating on the water surface is considered from a historical point of view, with specific emphasis on the development of a theoretical framework and equations of motion for long waves in shallow water. This provides the foundation for a subsequent discussion about tsunami wave propagation and run-up on a sloping beach, and in particular the role of wave dispersion for this problem. Wave tank experiments show that wave dispersion can play a significant role for the propagation and wave transformation of wave signals that include some higher frequency components. However, the maximum run-up height is less sensitive to dispersive effects, suggesting that run-up height can be adequately calculated by use of non-dispersive model equations.
\end{abstract}

\tableofcontents

\section{Introduction}

Surface gravity waves propagating on the air-sea interface are categorized as \emph{long waves} when their wave length $\lambda$ is large compared with the average water depth $h$ in a water basin. Waves of this type include very large scale phenomena such as tides and seiches, and local phenomena such as non-breaking shoaling waves at the coast. In recent years much attention has been connected with the study of tsunamis, with respect to their propagation over vast distances in open ocean, their transformation in coastal waters and their inundation of coastal areas. A range of different model equations have been discussed in connection with long wave propagation, including nonlinear shallow water equations (NLSW) and Boussinesq-type equations, which differ in their ability to represent nonlinear and dispersive effects. While elaborate model equations may provide more accurate representation of the wave propagation and transformation, they are generally more computationally demanding to integrate over time. In practical cases where a prediction of the wave behaviour is needed quickly, such as for a tsunami warning system, it has therefore been common practice to rely on simple NLSW equations rather than Boussinesq-type equations. Questions regarding the trade off between accuracy of prediction and efficiency of computation for shallow water model equations remain an active area of research to this day.

In this article we consider the problem of shallow water waves in a historical context, introducing some basic concepts of wave propagation. Thereafter we discuss the importance of these factors in the context of tsunami wave propagation and run-up on a sloping beach. Finally we consider some examples of different wave types, and assess the suitability of NLSW equations and a Boussinesq-type equation for each of these.


\section{Historical background}

The description of surface gravity wave propagation at the air-sea interface is one of the truly classical subjects in fluid mechanics, and developed in multiple stages with early contributions from some of the most prominent figures in science history such as Newton, Euler, Laplace, Lagrange, Poisson, Cauchy and Airy (see \cite{Darrigol2003,Craik2004,Craik2005} for historical references). For instance, in 1786 Lagrange demonstrated that small-amplitude waves would propagate in shallow water with a velocity of $c = \sqrt{gh}$, where $g$ represents the acceleration of gravity, and $h$ represents the water depth. Laplace (1776) was the first to pose the general initial-value problem for water wave motion, i.e. \emph{given a localized initial disturbance of the sea surface, what is the subsequent motion?} He was also the first to derive the full linear dispersion relation for water waves. A complete linear wave theory, which included wave dispersion, was later published by \cite{Airy1845}. Despite the long and extensive history of investigations into this problem, the study of dispersive surface gravity waves continues to be an active field of research to this day.


\subsection{Airy wave theory}

\begin{figure}
    \centering
    \includegraphics[width=\textwidth]{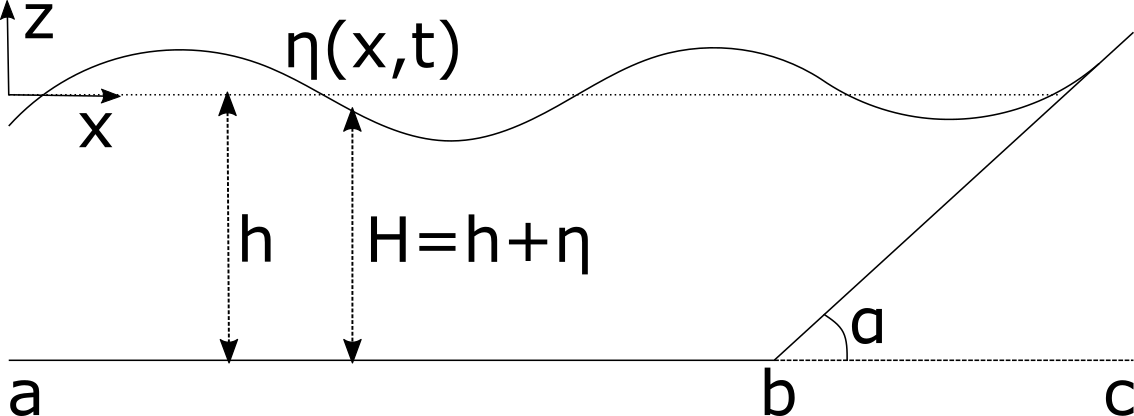}
    \caption{Reference coordinate system for surface gravity waves.}
    \label{fig:coordinate_system}
\end{figure}

In the following discussion we will restrict our attention to wave propagation in one horizontal dimension $x$ on a surface that can be displaced in the vertical $z$ direction (see Fig. \ref{fig:coordinate_system}). It is fairly simple to extend the theory to two horizontal dimensions, but we will not consider any examples where this is necessary, e.g. crossing wave patterns. A more thorough description of these equations can be found in standard fluid mechanics textbooks (see e.g. \cite{Kundu1990}).

A simple model equation for the propagation of surface gravity waves can be derived under the assumption of irrotational fluid motion, ignoring viscous effects, in which case the flow velocity components can be expressed in terms of a \emph{velocity potential} $\phi$, defined by
\begin{equation}
    u = \frac{\partial \phi}{\partial x} \qquad \textrm{and} \qquad w = \frac{\partial \phi}{\partial z} \, ,
\end{equation}
for horizontal and vertical velocity components $u$ and $w$, respectively. We do not consider effects due to surface tension, which is an important effect for short and steep waves but do not contribute significantly to long crested waves. Lastly, we assume that the water depth does not change very abruptly, i.e. that the water depth is fairly constant over the wave length.

The basic equation of motion is derived from the continuity equation
\begin{equation}
    \frac{\partial u}{\partial x} + \frac{\partial w}{\partial z} = 0 \, ,
\end{equation}
which is transformed to the Laplace equation
\begin{equation}
    \label{eq:Laplace}
    \frac{\partial ^2 \phi}{\partial x^2} + \frac{\partial ^2 \phi}{\partial z^2} = 0
\end{equation}
with substitution of the velocity potential. The sea bed is traditionally considered to be rigid and non-permeable, which implies that flow is only permitted along the bed profile. Under this assumption the boundary condition at the sea bed requires a zero normal velocity to the bed surface itself, which for a flat sea bed simplifies to 
\begin{equation}
    \label{eq:bottom_BC}
    w = \frac{\partial \phi}{\partial z} = 0 \qquad \textrm{at} \qquad z = -h \, .
\end{equation}
At the sea surface the fluid parcels are not restricted by any rigid boundary, and in fact the location of this free surface boundary is a variable we wish to determine by solving the equations of motion. If the wave field is sufficiently smooth we can assume that this boundary is well represented by a material surface, i.e. that fluid parcels at the boundary never leave the surface. Under this assumption the kinematic boundary condition prescribed at $z = \eta$ becomes
\begin{equation}
    \label{eq:kinematic_BC_full}
    \frac{\partial \eta}{\partial t} + 
    \left. u \frac{\partial \eta}{\partial x} \right|_{z=\eta} = 
    \left. \frac{\partial \phi}{\partial z} \right|_{z=\eta} \, .
\end{equation}
Provided the nonlinear term in eq. \eqref{eq:kinematic_BC_full} is sufficiently small, this expression can be replaced by the linear equation
\begin{equation}
    \label{eq:kinematic_BC_linear}
    \frac{\partial \eta}{\partial t} = \frac{\partial \phi}{\partial z} \qquad \textrm{at} \qquad z = \eta \, .
\end{equation}
Finally, the dynamic boundary condition prescribes that the pressure is continuous across the free surface boundary, which is expressed by the linear form of the Bernoulli equation
\begin{equation}
    \frac{\partial \phi}{\partial t} + \frac{P}{\rho} + gz = 0 \, ,
\end{equation}
where $P$ is pressure and $\rho$ is density. Assuming that the ambient pressure is zero at the free surface, i.e. $P = 0$ at $z = \eta$, this condition simplifies to
\begin{equation}
    \label{eq:dynamic_BC}
    \frac{\partial \phi}{\partial t} + g\eta = 0 \qquad \textrm{at} \qquad z = \eta \, .
\end{equation}
The complete boundary value problem is defined by eqs. \eqref{eq:Laplace}, \eqref{eq:bottom_BC}, \eqref{eq:kinematic_BC_linear} and \eqref{eq:dynamic_BC}.

In order to solve the equations we need to assume an initial wave form. By Fourier analysis it is possible to decompose any continuous disturbance into a sum of sinusoidal components, hence we will assume an initial condition specified by one such component with wave number $k$ and angular frequency $\omega$ 
\begin{equation}
    \label{eq:sinusoidal_solution}
    \eta(x,t) = a \cos(kx - \omega t) \, .
\end{equation}
The wave number and angular frequency represents the number of wave cycles (in radians) per unit length and unit time, respectively, and are defined in terms of the wave length $\lambda$ and wave period $T$ as
\begin{displaymath}
    k = \frac{2\pi}{\lambda} \qquad \textrm{and} \qquad \omega = \frac{2\pi}{T} \, .
\end{displaymath}
Solving the Laplace equation \eqref{eq:Laplace} with kinematic boundary conditions \eqref{eq:bottom_BC} and \eqref{eq:kinematic_BC_linear} with this initial condition results in a velocity potential
\begin{equation}
    \phi = \frac{a\omega}{k} \frac{\cosh k(z+h)}{\sinh kh} \sin(kx - \omega t) \, ,
\end{equation}
and the original velocity components become
\begin{equation}
    u = a\omega \frac{\cosh k(z+h)}{\sinh kh} \cos(kx - \omega t) \, ,
\end{equation}
\begin{equation}
    w = a\omega \frac{\sinh k(z+h)}{\sinh kh} \sin(kx - \omega t) \, .
\end{equation}
By combining this solution with the dynamic boundary condition \eqref{eq:dynamic_BC} we find the \emph{dispersion relation} between $k$ and $\omega$ as
\begin{equation}
    \label{eq:linear_dispersion}
    \omega = \sqrt{gk \tanh kh} \, ,
\end{equation}
and the corresponding \emph{phase velocity}
\begin{equation}
    \label{eq:linear_phase_speed}
    c_p \equiv \frac{\omega}{k} = \sqrt{\frac{g}{k} \tanh kh} \, .
\end{equation}

The dispersion relation eq. \eqref{eq:linear_dispersion} links the wave number $k$ with the angular frequency $\omega$, and represents a necessary condition for consistency of linear wave solutions for the equations of motion. This implies that sinusoidal wave solutions eq. \eqref{eq:sinusoidal_solution} can exist if and only if the wave length and period are strictly linked with each other according to eq. \eqref{eq:linear_dispersion}. Eq. \eqref{eq:linear_phase_speed} demonstrates that the speed of propagation for linear waves depend on the wave number (equivalently, wave length), hence an initial disturbance that contains wave components with various wave numbers will tend to separate into clusters of individual components as the waves propagate away from the source.

For waves in a dispersive medium, the energy of wave components does not propagate with the phase velocity $c_p$ eq. \eqref{eq:linear_phase_speed}, but with the \emph{group velocity} $c_g = d \omega / dk$. With the dispersion relation defined in eq. \eqref{eq:linear_dispersion}, the group velocity becomes
\begin{equation}
    \label{eq:linear_group_speed}
    c_g \equiv \frac{d \omega}{ dk} = \frac{c_p}{2} 
    \left[ 1 + \frac{2kh}{\sinh 2kh} \right] \, .
\end{equation}


\subsubsection{Deep and shallow water approximations}

As seen in eq. \eqref{eq:linear_phase_speed}, the phase velocity for wave components with different wave numbers depend on the hyperbolic tangent function. It is instructive to consider the behaviour of this equation in the deep and shallow water conditions, which is determined by the value of $kh$ (i.e. water depth relative to wave length). A commonly used classification is to consider \mbox{$kh \geq \pi$} as deep water, \mbox{$kh \leq \pi/10$} as shallow water, and \mbox{$\pi/10 < kh < \pi$} as intermediate water depth. This classification should be considered a ``rule-of-thumb'' rather than a strict rule, as the dispersion relation varies continuously over the range of $kh$.

For the deep water approximation, the hyperbolic functions in eqs. \eqref{eq:linear_phase_speed} and \eqref{eq:linear_group_speed} can be approximated as
\begin{displaymath}
    \lim_{kh \rightarrow \infty} \tanh kh = 1 \qquad \textrm{and} \qquad \lim_{kh \rightarrow \infty} \frac{2kh}{\sinh 2kh} = 0 \, ,
\end{displaymath}
in which case the phase and group speed in deep water becomes 
\begin{equation}
    \label{eq:deep_water_speed}
    c_p = \sqrt{\frac{g}{k}} \qquad \textrm{and} \qquad c_g = \frac{c_p}{2} \, ,
\end{equation}
respectively. This implies that short waves in deep water propagate slower than longer waves, and the wave energy propagate slower than the wave phase. Note that eq. \eqref{eq:deep_water_speed} is derived under the assumption that gravity is the only relevant restoring force, which is not always correct. For instance, at very short wave lengths (cm scale at the air-water interface) surface tension becomes the dominant restoring force, which allows shorter wave components to propagate faster than longer wave components. The velocity components simplify to
\begin{equation}
    u = a\omega e^{kz} \cos (kx - \omega t) \, ,
\end{equation}
\begin{equation}
    w = a\omega e^{kz} \sin (kx - \omega t) \, ,
\end{equation}
which are circular orbits with a radius of $a$ at the surface. It should be noted that the linear wave theory assumes that effects due to the finite wave amplitude are negligible. In reality waves have a finite amplitude, which induces a slow drift in the direction of wave propagation, and therefore the orbits of fluid parcels are not perfect circles but display a coil-like behaviour. This effect is called \emph{Stokes drift}.

In shallow water the hyperbolic functions in eq. \eqref{eq:linear_phase_speed} can be replaced by $kh$ because
\begin{displaymath}
    \tanh kh = kh + \mathcal{O} \left( (kh)^2 \right) \, , \quad \textrm{as} \quad kh \rightarrow 0 \, ,
\end{displaymath}
and the hyperbolic function in eq. \eqref{eq:linear_group_speed} can be approximated as
\begin{displaymath}
    \lim_{kh \rightarrow 0} \frac{2kh}{\sinh 2kh} = 1 \, ,
\end{displaymath}
hence the simplified expressions for the phase speed and group speed become
\begin{equation}
    c_p = \sqrt{gh} \qquad \textrm{and} \qquad c_g = c_p \, ,
\end{equation}
respectively. In this case the phase velocity is not dependent on the wave number $k$, hence waves in the shallow water limit are non-dispersive. This is also reflected in the group velocity, which becomes identical to the phase velocity in shallow water. In the special case of unidirectional flow, the shallow water wave field therefore becomes stationary in the coordinate system that follows the phase speed $c_p$. The velocity components for shallow water waves (of small but finite depth) are
\begin{equation}
    u = \frac{a\omega}{kh} \cos (kx - \omega t) \, ,
\end{equation}
\begin{equation}
    w = a\omega \left( 1 + \frac{z}{h} \right) \sin (kx - \omega t) \, .
\end{equation}
These are elliptic orbits where the vertical component is much smaller than the horizontal.


\subsection{Nonlinear long waves}

The wave theory developed by Airy is a linear system, requiring both the underlying equations of motion and boundary conditions to be linear, and therefore any wave solution to this system must conform to the superposition principle. This means that the net response to the system of two or more stimuli can be established by determining the response of each stimulus separately, and subsequently adding these together. Equivalently, any linear combination or scaling of valid solutions will produce a new valid solution to the problem. In particular, this means that the wave amplitude, which can be altered by a scalar multiplication, must be an independent variable that can not have any functional dependence to other wave properties. This property is specific for linear systems, whereas for nonlinear systems the wave amplitude will normally be linked with other wave properties. In fact, waves of this type had already been described at the time when Airy published his account.

A few years prior to the publication of Airy's wave theory, the naval engineer John Scott Russell had published accounts of observations and experiments devoted to surface gravity waves \citep{Russell1844}. Russell seem to have devoted most of his efforts to explain wave generation and propagation in channels, which was of practical importance for inland waterway transport at that time. A particularly famous account describes his first observation of a \emph{large, solitary, progressive wave}, which was generated by a boat in a channel and proceeded to propagate upstream of the boat. Russel was able to follow this wave on horseback for more than a mile, and while it retained its original shape it then gradually subsided. In a series of subsequent experiments he determined that the wave progressed upstream with a velocity $c = \sqrt{g(h+\eta)}$, and that the wave making resistance against the boat motion was at a maximum when the boat was traveling at this speed. He also proposed that tidal motion could be explained as solitary waves of very large extent, and suggested a mechanism whereby the tidal motion could generate tidal bores in rivers and channels.

Airy devoted some attention to Russell's experiments, but he dismissed Russell's treatment of solitary waves. According to Airy's wave theory, maintaining such a singular disturbance in the absence of any additional force would require the surface slope of the disturbance to be constant, but since the slope should vanish at infinity such a disturbance could not exist. The existence and importance of solitary waves remained a contested issue for several decades after the initial treatments by Russell and Airy. For example, the prominent scientist Georges Gabriel Stokes first dismissed the possibility of such waves and their relevance to tidal motion in his 1846 hydrodynamic researches review \citep{Stokes1846}, but later became supportive of the idea after researching finite oscillatory waves. 

In 1870 Adhémar Jean Claude Barré de Saint-Venant published an account of tidal bores in rivers (named \emph{mascaret} in French), and the following year \citep{Saint-Venant1871} he presented a set of equations that described the phenomenon
\begin{equation}
    \label{eq:saint-venant_cont}
    \frac{\partial A}{\partial t} + \frac{\partial (Au)}{\partial x} = 0 \, ,
\end{equation}
\begin{equation}
    \label{eq:saint-venant_moment}
    \frac{\partial u}{\partial t} + u\frac{\partial u}{\partial x} + g\frac{\partial \eta}{\partial x} = - \frac{P_{w}}{A} \frac{\tau}{\rho} \, ,
\end{equation}
where $A(x,t)$ is the channel cross-section area, $u(x,t)$ is the depth-averaged horizontal velocity component, $P_{w}(x,t)$ is the length of wetted channel perimeter at the cross-section, $\tau(x,t)$ is the wall shear stress, and $\rho$ is the water density. The set of equations \eqref{eq:saint-venant_cont} and \eqref{eq:saint-venant_moment} represent conservation of mass and balance of momentum, respectively, and is possibly the first version of NLSW equations to be presented in a publication. Due to the friction force induced by the shear stress at channel walls, the momentum of an initial disturbance will not be conserved in the model system. The shallow water equations (\ref{eq:saint-venant_cont},~\ref{eq:saint-venant_moment}) can describe the propagation of a solitary wave, but the wave will transform over time, with a steepening of the wave front and a decrease in the slope behind the crest. While this behaviour nicely described the transformation of a regular tidal wave to a tidal bore in a channel, it did not provide an adequate framework for describing the solitary waves of constant shape observed by Russell. Although similar shallow water equations had been presented prior to Saint-Venant's treatment of mascarets \citep{Fenton2010}, the one-dimensional (1D) version of the shallow water equations are often referred to as \emph{Saint-Venant equations} in honor of his contribution to understand shallow water hydrodynamics.

The same year as Saint-Venant presented the NLSW equations for description of mascarets, one of his disciples,  Joseph Boussinesq, presented the first approximate solution of a solitary wave propagating without deformation \citep{Boussinesq1871}, which finally provided a firm theoretical support for the existence of Russell's wave. The following year \citep{Boussinesq1872} he presented a derivation of equations which permitted his wave solution
\begin{equation}
     \label{eq:boussinesq_cont}
    \frac{\partial \eta}{\partial t} + \frac{\partial (H u_b)}{\partial x} = \frac{h^3}{6} \frac{\partial ^3 u_b}{\partial x^3} \, ,
\end{equation}
\begin{equation}
    \label{eq:boussinesq_moment}
    \frac{\partial u_b}{\partial t} + u_b \frac{\partial u_b}{\partial x} + g \frac{\partial \eta}{\partial x} = \frac{h^2}{2} \frac{\partial ^3 u_b}{\partial t \partial x^2} \, ,
\end{equation}
where $H = h + \eta$ (see Fig. \ref{fig:coordinate_system}) and $u_b$ is the horizontal velocity at the sea bed $z = -h$. This is the original version of what is now called \emph{Boussinesq equations}. In the absence of higher-order derivatives (right-hand side of eqs. (\ref{eq:boussinesq_cont},\ref{eq:boussinesq_moment})) the Boussinesq system becomes equivalent to the Saint-Venant equations (\ref{eq:saint-venant_cont},\ref{eq:saint-venant_moment}) without a friction term. Boussinesq derived his equations from the Euler equations by eliminating the explicit dependence on the vertical coordinate $z$ in these equations, while retaining nonlinear terms of highest order. This procedure, which is now commonly used when deriving shallow water equations, does not \emph{a priori} stipulate the vertical reference level to be used for the horizontal velocity component or which higher-order terms to retain in the derivation. Numerous variations of Boussinesq-type systems can therefore be derived by selecting different reference variables and forms of nonlinear terms, resulting in equations with slightly different dispersive and nonlinear properties, as well as numerical stability properties. A particularly useful variation was derived by \cite{Peregrine1967}
\begin{equation}
    \label{eq:peregrine_cont}
    \frac{\partial \eta}{\partial t} + \frac{\partial (H u)}{\partial x} = 0 \, ,
\end{equation}
\begin{equation}
    \label{eq:peregrine_moment}
    \frac{\partial u}{\partial t} + u \frac{\partial u}{\partial x} + g \frac{\partial \eta}{\partial x} - \frac{h}{2} \frac{\partial ^3(hu)}{\partial ^2x \partial t} + \frac{h^2}{6} \frac{\partial ^3 u}{\partial x^2 \partial t} = 0 \, ,
\end{equation}
which can be applied under gently varying depth conditions. While the achievement of Boussinesq is widely recognized at present time, his results were not immediately seized upon by his contemporaries. Five years after Boussinesq presented his solitary wave solution, Lord Rayleigh independently derived a long wave equation for the solitary wave of constant shape \citep{Rayleigh1876}. When Korteweg and de Vries later derived their famous \emph{KdV equation}, they reference to Rayleigh's work but were apparently unaware of the earlier contribution by Boussinesq \citep{Korteweg1895}.


\subsection{Model equations for long wave run-up on a beach}

In the classical formulations of long wave equations it is usually assumed that the waves propagate in a water basin with small and gentle changes in water depth. However, we would like to apply these model equations to study wave run-up on a beach, and this requires some modifications to the standard equation formulations. In the following we consider a depth profile
\begin{equation}
    \label{eq:depth_profile}
    h(x) = 
    \begin{cases}
    h_0 \, , & \textrm{ if } x \in [a,b] \\
    h_0 - (x - b) \tan \alpha \, , & \textrm{ if } x \in [b,c]
    \end{cases} \, ,
\end{equation}
with waves approaching the beach from the offshore point $a$ (see Fig. \ref{fig:coordinate_system}). The modified NLSW equations are defined as
\begin{equation}
    \label{eq:nlsw_cont}
    \frac{\partial H}{\partial t} + \frac{\partial (Hu)}{\partial x} = 0 \, ,
\end{equation}
\begin{equation}
    \label{eq:nlsw_moment}
    \frac{\partial (Hu)}{\partial t} + \frac{\partial}{\partial x} 
    \left( Hu^2 + \frac{g}{2}H^2 \right) = gH \frac{\partial h}{\partial x} \, , 
\end{equation}
where $u(x,t)$ is the depth-averaged flow velocity. For comparison, we use a Boussinesq-type equation based on Peregrine's formulation, which we call the modified Peregrine equations
\begin{equation}
    \label{eq:mPer_cont}
    \frac{\partial H}{\partial t} + \frac{\partial Q}{\partial x} = 0 \, ,
\end{equation}
\begin{gather}
    \label{eq:mPer_moment}
    \left( 1 + \frac{1}{3} \frac{\partial H^2}{\partial x} - \frac{H}{6} \frac{\partial ^2 H}{\partial x^2} \right) \frac{\partial Q}{\partial t} - \frac{H^2}{3} \frac{\partial ^3 Q}{\partial x^2 \partial t}  \\
    - \frac{H}{3} \frac{\partial H}{\partial x} \frac{\partial ^2Q}{\partial x \partial t} \nonumber + \frac{\partial}{\partial x} \left( \frac{Q^2}{H} + \frac{g}{2} H^2 \right) =  gH \frac{\partial h}{\partial x} \, ,
\end{gather}
where $Q = Hu$ represent the horizontal momentum. The modified Peregrine equations have been studied in detail in \cite{Duran2018}.


\subsection{Numerical method}

In the following discussion we will apply numerical methods for integration of the Boussinesq equations over time. For simple channel geometries it is possible to derive exact solitary wave and periodic wave solutions to the Boussinesq equations 
\citep{Clarkson1990,Chen1998,Yan1999}. Furthermore, the run-up properties of such general wave solutions can be investigated by analytical methods for some regular beach profiles \citep{Pelinovsky_Mazova_1992_NatHaz,Didenkulova_etal_2007_Springer,Didenkulova2011}. However, such analytical methods are not practical when considering general wave types and variable depth conditions. The numerical model we use is based on a finite volume method for both the modified NLSW and modified Peregrine equations \citep{Dutykh2011,Duran2018}. This involves discretization of the of the governing equations, and obtaining solutions on a finite mesh covering the model domain. In the finite volume method, the divergence theorem is applied to convert divergence terms in the differential equations to surface integrals, which are evaluated as fluxes at the cell surfaces in the mesh. Finite volume methods are particularly useful for problems where quantities should be preserved, e.g. mass or momentum, since whatever quantity flows out of one grid cell surface will be identical to the inflow of the neighbouring grid cell. 

The simplest approximation to a solution in the finite volume formulation is obtained by considering all variables as constant within each grid cell, whereby a piece-wise constant solution can be obtained. However, by this approach the spatial discretization error will be determined by the grid size. In order to obtain more accurate results, a common method is to replace the piece-wise constant data with a piece-wise polynomial representation of the solution. In our simulations we have applied the non-oscillatory UNO2 scheme, which is designed to constrain the number of local extrema in the numerical solution at each time step \citep{Harten1987}.

Integration of the solution forward in time is achieved by the Bogacki-Shampine time stepping method \citep{Bogacki1989}. This is a version of a Runge-Kutta method, and is a third order method with four stages. An embedded second order method is used to estimate the local error and if necessary adapt the time-step size.


\section{Tsunami propagation and run-up}

Developing model equations that adequately describe the propagation
and run-up of tsunamis is a challenging task. Suggested model formulations range from simple nonlinear shallow water theory (NLSW) to the very elaborate fully nonlinear Navier-Stokes theory (FNS), with Boussinesq theory occupying an intermediate place in between. NLSW has often been favored for long wave run-up calculations over dispersive wave models represented by Boussinesq-type approximations. Wave run-up calculated using dispersive model formulations is prone to numerical instabilities, which make computations more sensitive to numerical parameters \citep{Bellotti2002}. Furthermore, the Boussinesq terms in the dispersive model tend to zero at the shoreline, so that dispersive equations simplify to NLSW in this region \citep{Madsen1997}. 

High accuracy can be achieved by applying the fully nonlinear Navier-Stokes equations, but this approach requires large computational resources and a lengthy integration time, making it unsuitable for operational forecasting in ocean-wide or even regional scale applications. \cite{Horrilo2006} studied dispersive effects during 2004 Indian Ocean tsunami propagation by comparing NLSW with the fully nonlinear Navier-Stokes equations (FNS). They concluded that NLSW offered the more suitable framework for hazard assessments, providing an adequate assessment at a very low computational cost. Although the NLSW model tended to over-predict the maximum wave run-up, the over-prediction was considered to be within a reasonable range for a safety buffer, and hence did not degrade the overall assessment. For tsunami warning purposes it is of critical importance to determine the time of arrival of the leading wave to different coastal sections. These leading waves are usually well described by NLSW, whereas the trailing wave train may contain shorter wave components that are more sensitive to wave dispersion \citep{Lovholt2012}. For this reason, the NLSW is often considered to be more appropriate than more elaborate Boussinesq-type methods for warning purposes \citep{Glimsdal2013}. Note, that maximum wave is often not the first one, at least for tsunamis propagating over a long distance, see, for example, \cite{Candella2008}.


\subsection{Wave tank experiment}

Wave tank experiments were carried out at the Large Wave Flume (GWK) located in Hannover, Germany, and is the world's largest publicly available research facility of its kind. It has a length of about 310 m usable for experiments, a width of 5 m, and a maximum depth of 7 m. Access to this facility was  granted by the Integrating Activity HYDRALAB IV program, and was carried out over two periods; 10--16 Oct. 2012 and 29 July -- 9 Aug. 2013. The basic experiment setup consisted of a wave generator at one end of the flume, a 251 m channel of constant depth, and a ramp of 1:6 slope at the opposite end of the flume representing the beach. The water depth in the channel was kept at a constant $h_0 = 3.5$ m for all the experiments. Wave gauges were placed at 16-18 locations along the channel to measure the waves propagating in the channel and up the slope. The wave run-up was measured by a capacitance probe and also recorded by two regular video cameras. A series of experiment runs were performed with different initial wave signals, and with varying roughness of the ramp slope surface. Details of the experiments are described in \cite{Didenkulova2013c}.


\section{Measured and modeled wave propagation and run-up}

\begin{table}[htbp]
    \centering
    \begin{tabular}{lccc}
    \hline \\
    Type of waves & Wave       & Initial wave  & Experimental \\
                  & period (s) & amplitude (m) & run-up (m)   \\
    \hline
    Sine wave        & 20 & 0.20 & 0.571 \\
    Bi-harmonic wave & 20 & 0.12 & 0.794 \\
    Wake-like train  & 20$\rightarrow$10 & $\sim$ 0.10 & 0.517 \\
    Positive pulse   & 20 & 0.15 & 0.438 \\
    \hline
    \end{tabular}
    \caption{Parameters for four experiment runs of different wave types, and the measured run-up for each case.}
    \label{tab:wave_types}
\end{table}

\begin{figure}
    \centering
    \begin{subfigure}[t]{0.9\textwidth}
        \includegraphics[width=\textwidth]{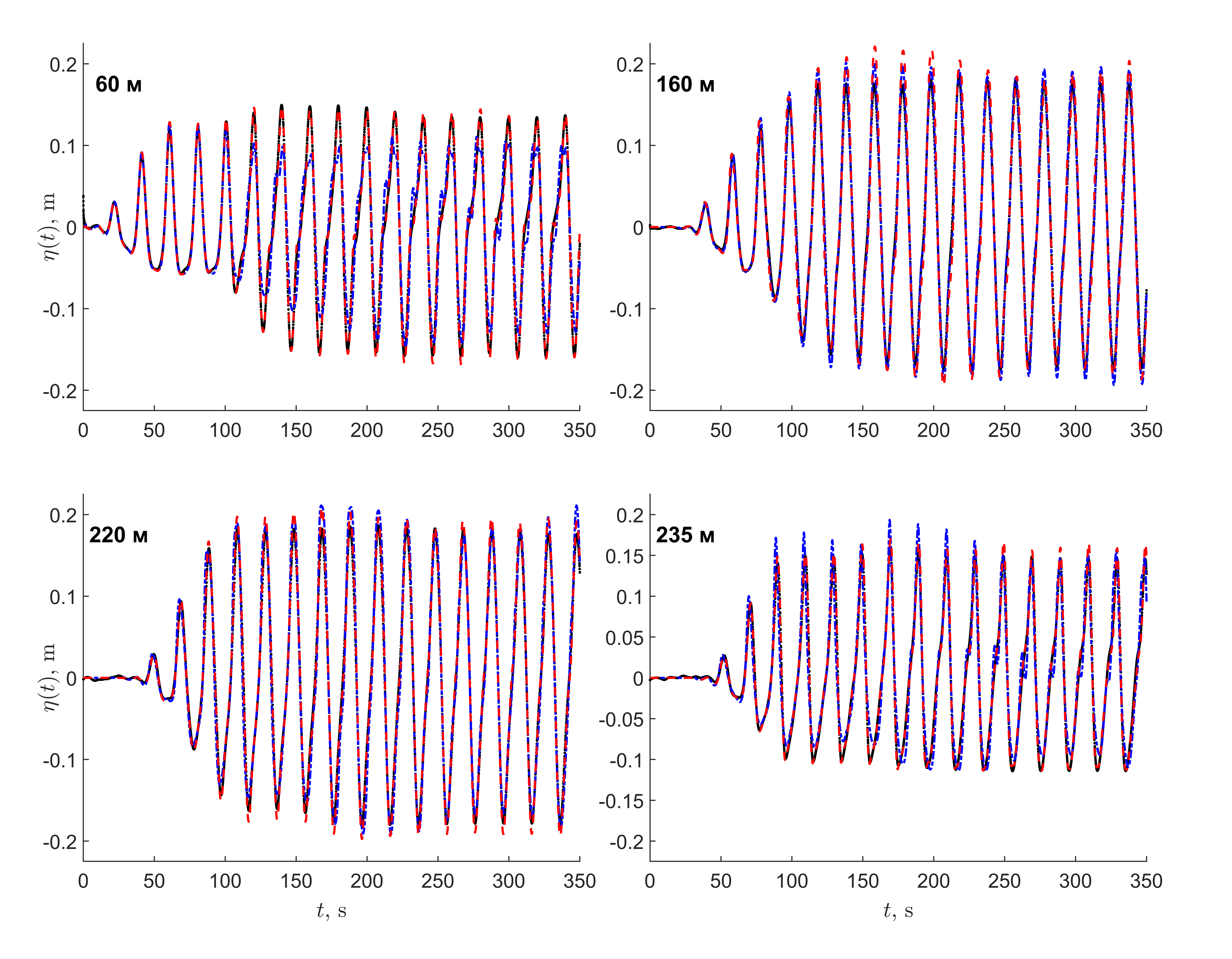}
        \caption{Water surface elevation at different wave gauges ($x$ = 60~m, 160~m, 220~m, and 235~m from the wave maker)}
        \label{fig:Fig2}
    \end{subfigure}
    \\
    \begin{subfigure}[t]{0.9\textwidth}
        \includegraphics[width=\textwidth]{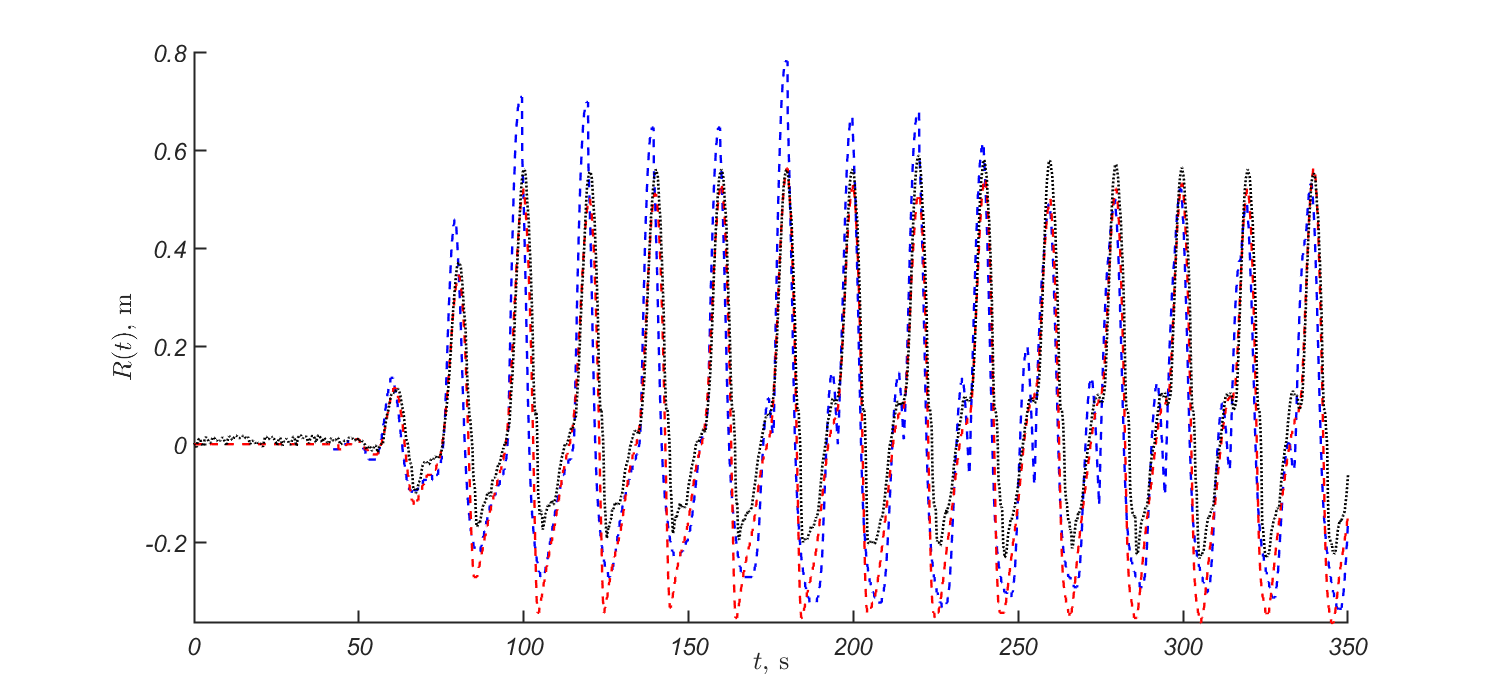}
        \caption{Run-up height}
        \label{fig:Fig3}
    \end{subfigure}
    \caption{Wave propagation and run-up for a sine wave with $A$ = 0.2 m and $T$ = 20 s on a beach slope $\tan \alpha$ = 1:6, mPer is shown with the red dash line, NLSW solution is shown with Blue dash-dots line and the experimental record is shown with the black dots line.}
    \label{fig:sine_wave}
\end{figure}

\begin{figure}
    \centering
    \begin{subfigure}[t]{0.9\textwidth}
        \includegraphics[width=\textwidth]{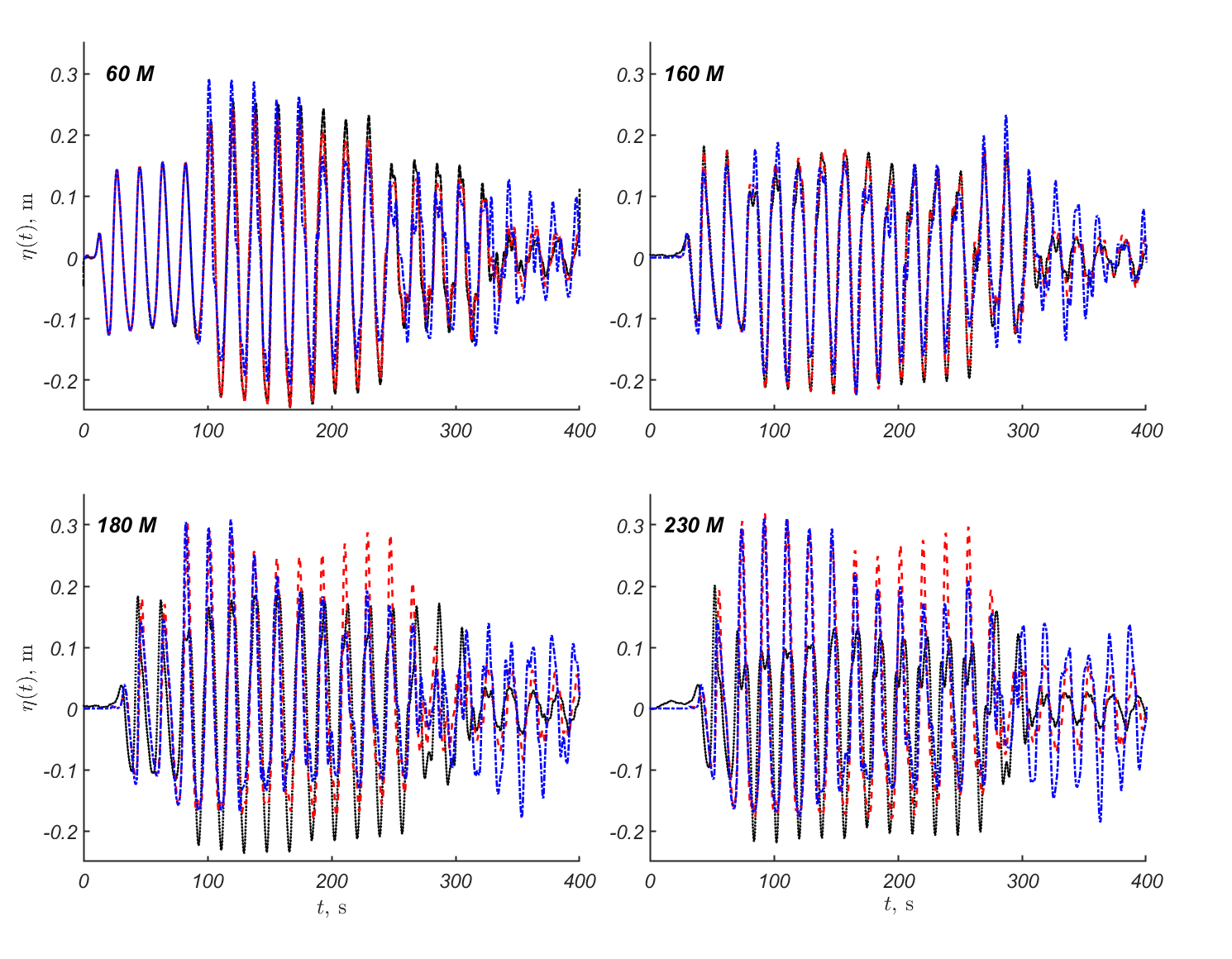}
        \caption{Water surface elevation at different wave gauges ($x$ = 60 m, 160 m, 180 m, and 230 m from the wave maker)}
        \label{fig:Fig4}
    \end{subfigure}
    \\
    \begin{subfigure}[t]{0.9\textwidth}
        \includegraphics[width=\textwidth]{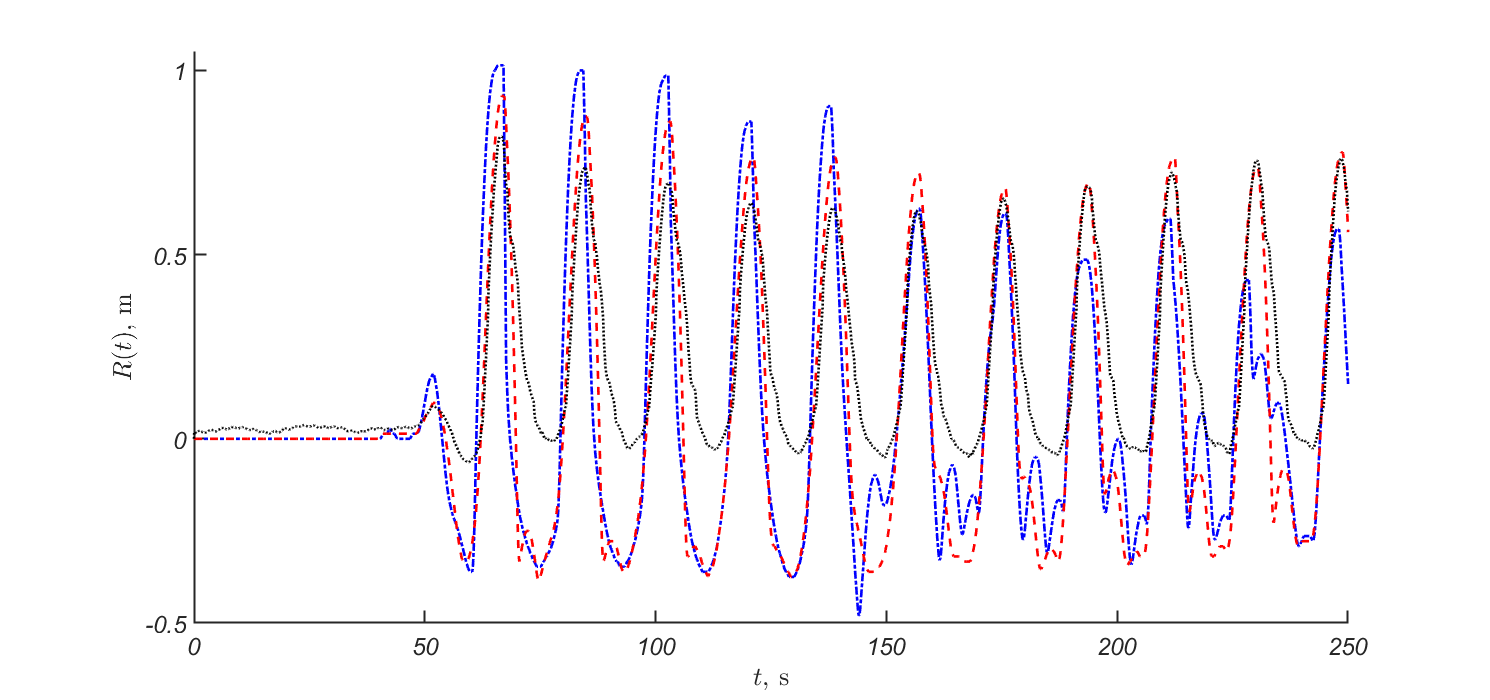}
        \caption{Run-up height}
        \label{fig:Fig5}
    \end{subfigure}
    \caption{Wave propagation and run-up for a bi-harmonic wave with $A$ = 0.12 m and $T$ = 20 s on a beach slope $\tan \alpha$ = 1:6, mPer is shown with the red dash line, NLSW solution is shown with Blue dash-dots line and the experimental record is shown with the black dots line.}
    \label{fig:bi-harmonic_wave}
\end{figure}

\begin{figure}
    \centering
    \begin{subfigure}[t]{0.9\textwidth}
        \includegraphics[width=\textwidth]{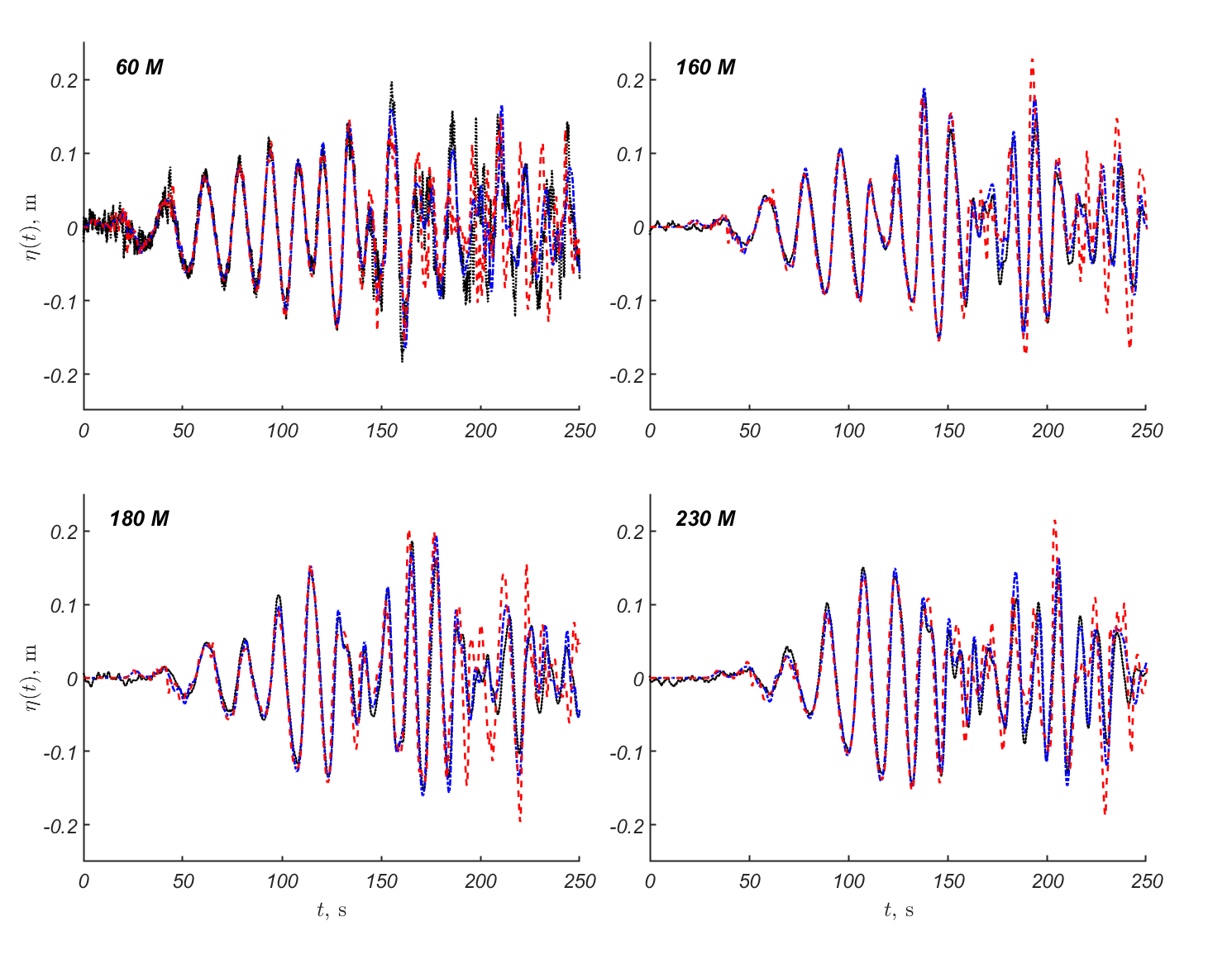}
        \caption{Water surface elevation at different wave gauges ($x$ = 60 m, 160 m, 180 m, and 230 m from the wave maker)}
        \label{fig:Fig6}
    \end{subfigure}
    \\
    \begin{subfigure}[t]{0.9\textwidth}
        \includegraphics[width=\textwidth]{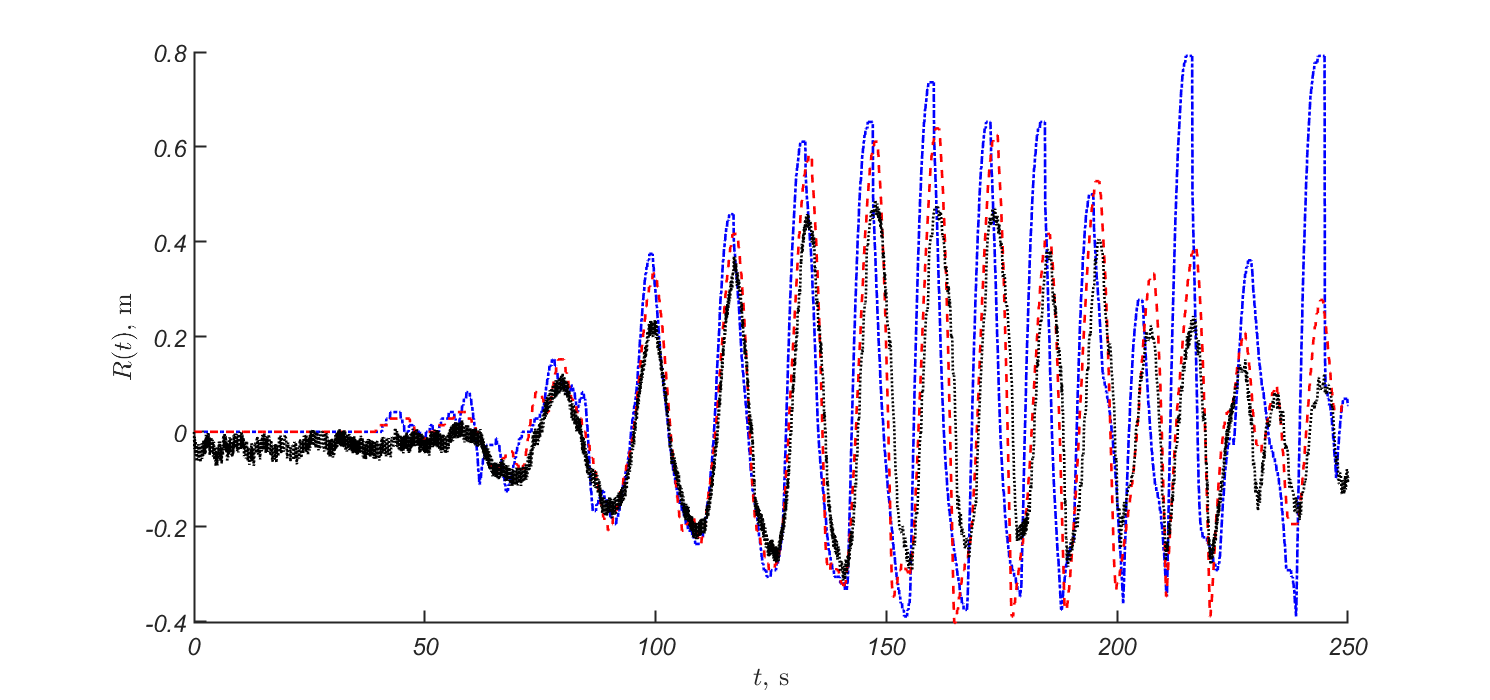}
        \caption{Run-up height}
        \label{fig:Fig7}
    \end{subfigure}
    \caption{Wave propagation and run-up for a wake-like wave train with $A$ = 0.1~m and $T \in [10,20]$~s on a beach slope $\tan \alpha$ = 1:6, mPer is shown with the red dash line, NLSW is shown with Blue dash-dots line and the experimental record is shown with the black dots line.}
    \label{fig:wake-like_wave}
\end{figure}

\begin{figure}
    \centering
    \begin{subfigure}[t]{0.9\textwidth}
        \includegraphics[width=\textwidth]{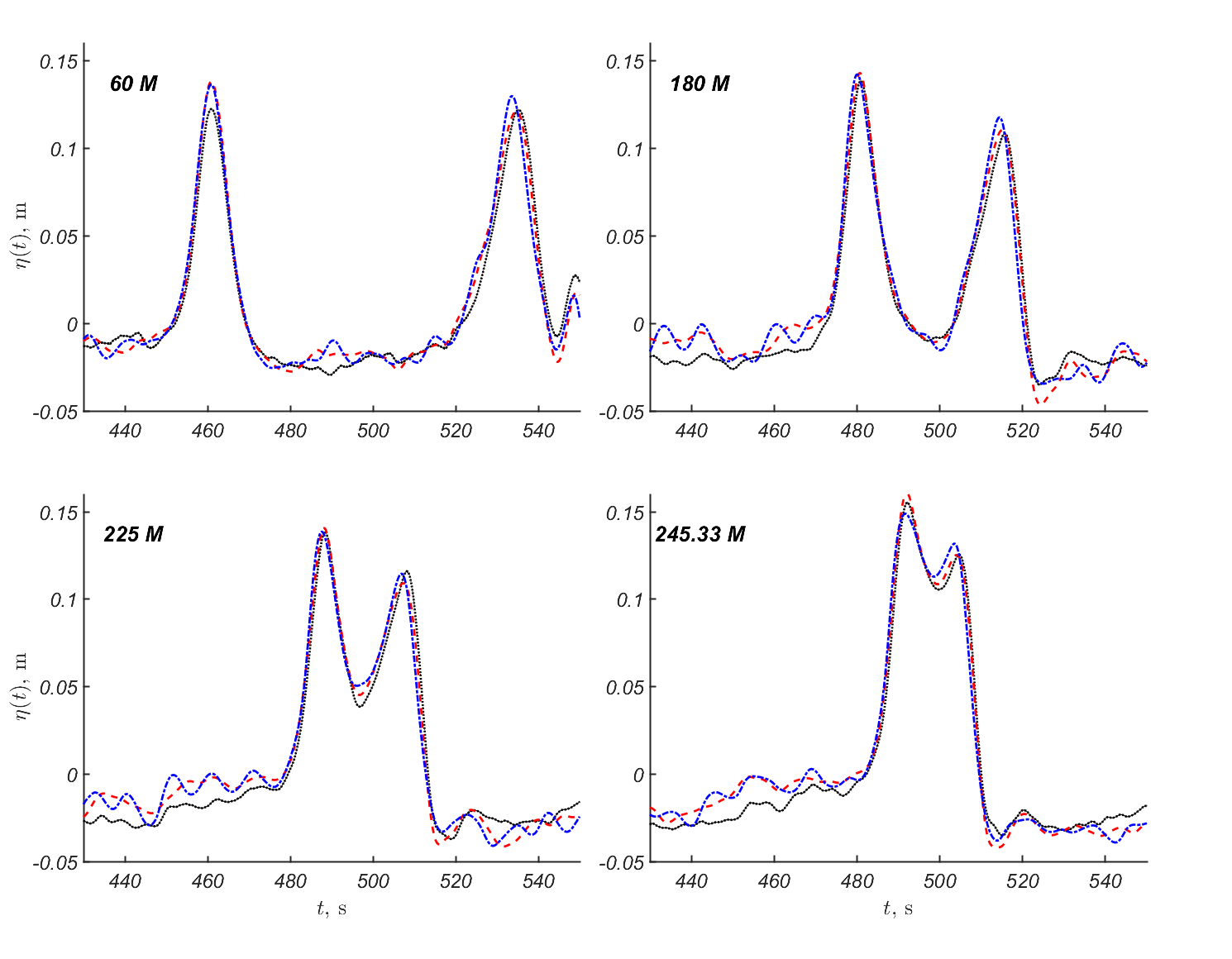}
        \caption{Water surface elevation of solitary wave at different wave gauges ($x$ = 60 m, 160 m, 2250 m, and 245.53 m from the wave maker)}
        \label{fig:Fig8}
    \end{subfigure}
    \\
    \begin{subfigure}[t]{0.9\textwidth}
        \includegraphics[width=\textwidth]{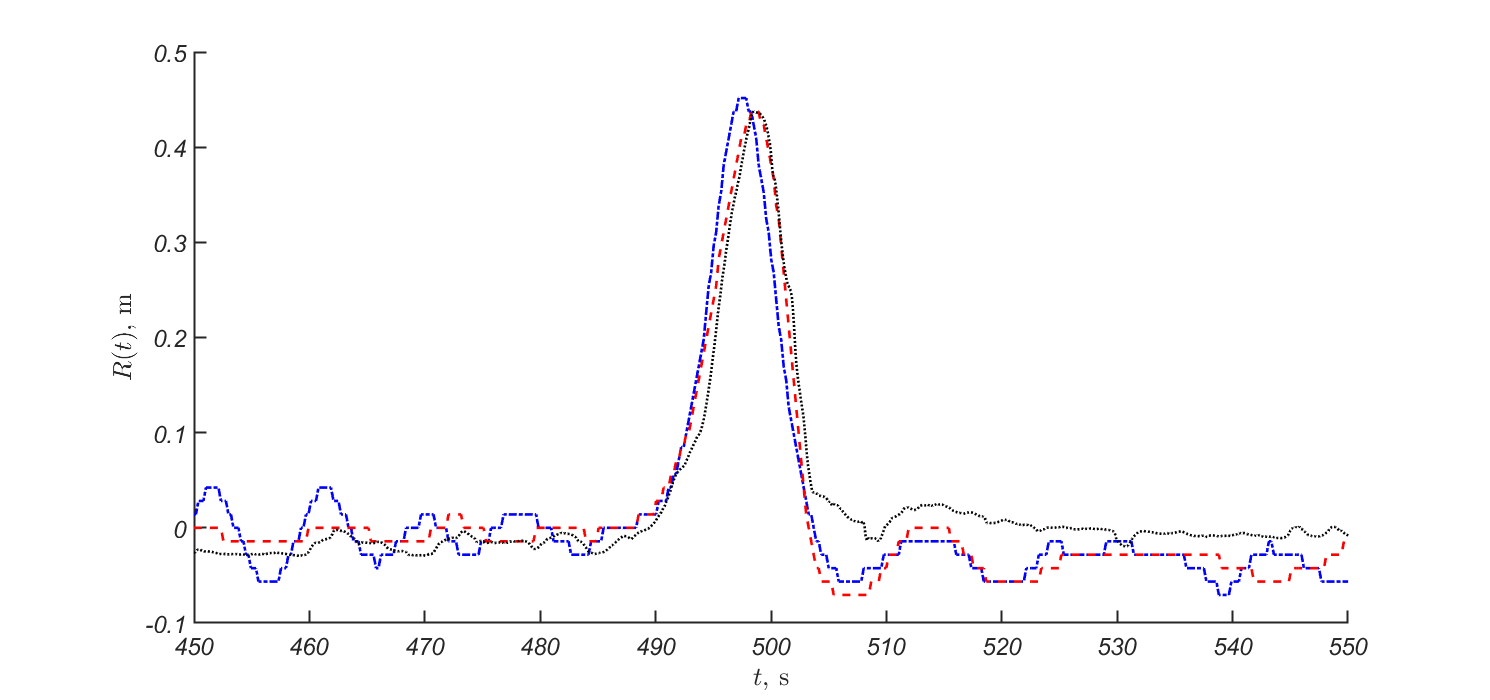}
        \caption{Run-up height}
        \label{fig:Fig9}
    \end{subfigure}
    \caption{Wave propagation and run-up for a single positive pulse (solitary wave) with $A$ = 0.15~m and $T \in [10,20]$~s on a beach slope $\tan \alpha$ = 1:6, mPer is shown with the red dash line, NLSW is shown with Blue dash-dots line and the experimental record is shown with the black dots line.}
    \label{fig:single_pulse}
\end{figure}

In order to illustrate the wave transformation and run-up properties for different wave signals, we consider the four experimental test cases listed in Table \ref{tab:wave_types}. These consist of a regular sine wave, a bi-harmonic wave signal, a wave train that resembles a ship wake signal, and a single positive pulse. The wave maker produced waves of period $T = 20$~s, which remained constant for the sine and bi-harmonic signals, but gradually reduced to $T=10$~s for the wake-like train. The initial wave amplitude was different for each experiment, with the largest initial wave amplitude $A = 0.20$~m for the sine wave (Fig. \ref{fig:Fig2}). However, the bi-harmonic wave signal contained two wave components with amplitude $A = 0.12$~m that could interfere constructively to produce instances of larger amplitude wave peaks than the sine wave (Fig. \ref{fig:Fig4}). The wake-like train did not contain waves of equal amplitude. Instead, initially long, low-amplitude waves were followed by progressively shorter and larger-amplitude waves (Fig. \ref{fig:Fig6}). The single positive pulses were generated with $A = 0.15$ m, but were not initiated as stable solitary wave shapes and hence reduced in amplitude to approximately $A = 0.10$~m at an early stage during the wave propagation (Fig. \ref{fig:Fig8}). Each figure shows a comparison between the experimental record and two model results; the dispersive modified Peregrine model (hereafter mPer) and the NLSW model solutions.

Figure \ref{fig:sine_wave} show the sine wave propagation and run-up. In this case mPer is fairly close to the measured waves throughout the propagation phase and for the run-up, although there is a tendency for mPer to underestimate the run-up height. It is noticeable that NLSW has a lower wave height near the wave maker than the measured wave, but increase in amplitude relative to the reference solution, and in the final stage produce significantly larger run-up values than the measured values. It is clear that the dispersive properties of mPer in this case balance the nonlinear effect to produce a relatively stable wave train, while this feature is missing for NLSW and therefore results in excessive nonlinear steepening and amplification. Note that the capacitance runup gauge does not record the wave form correctly in the receding phase. The reason is that the wires are submerged in the thin near-surface layer of water when the bulk of the wave is gone. Therefore, only the raising front phase of experimentally measured runup should be used for comparison with the simulations and the rundown values indicated by the gauge should be ignored.

Figure \ref{fig:bi-harmonic_wave} show the bi-harmonic wave propagation and run-up. In this case we again see a reasonably good agreement in wave structure between mPer and the measurements, but now there is a clear tendency that mPer underestimates the wave amplitude both in the propagation phase and the run-up phase. The NLSW solution looks fairly reasonable in the early stages, but significant discrepancies appear at $x = 180$~m and $x = 230$~m in the later stages of the wave train. The bi-harmonic signal is particularly sensitive to the phase speed as wave components may interfere both constructively and destructively at different stages, hence the inclusion of wave dispersion plays a significant role in this case.

Figure \ref{fig:wake-like_wave} show a wave train with a wake-like structure, with a distinct envelope shape created by an initial long, low-amplitude wave followed by shorter, higher amplitude waves. The initial phase of the wave train is captured well by both mPer and NLSW, although both models struggle to reproduce the later stages of the wave train. Both models also reproduce the run-up phase fairly well, although NLSW develops a slight phase shift relative to the reference solution, and both models severely over-estimates the run-up for the trailing waves.

Figure \ref{fig:single_pulse} show the wave propagation and run-up for single positive pulse waves. The model results for mPer and NLSW are remarkably similar for the propagation phase in this case, although both models tend to overestimate the wave amplitude slightly. This discrepancy can likely be explained by inaccuracies in the initial conditions for the wave, as it deviates slightly from a stable solitary wave form. The run-up results are likewise very similar between mPer, NLSW and the reference solution, but again we see the tendency that mPer underestimates the run-up height, whereas NLSW overestimates the run-up height and has a slight phase shift indicating that the propagation speed is slightly elevated relative to the reference solution.


\section{Concluding remarks}

The results in the preceding section demonstrate some of the capabilities of the NLSW and modified Peregrine equation systems for representation of long wave transformations. Both models compare well with the long single wave of positive polarity. For sine waves, bi-harmonic signals and dispersive wake-like signals the wave dispersion clearly plays a more prominent role, in which case NLSW does not adequately represent the high frequency components. Despite the differences in wave transformation and propagation, the differences in maximum wave run-up are quite modest, suggesting that the dispersive wave properties does not influence the resulting run-up to a significant extent. This would suggest that NLSW could be a suitable framework for prediction of tsunami events also in the future, despite the known shortcomings of the model equations for dispersive waves.

Research into surface gravity wave phenomena has a long and fascinating history. Modern day researchers benefit greatly by working within a framework where theories for e.g. Fourier analysis, ordinary and partial differential equations, potential theory, and perturbative methods, are well established. The emergence of computational resources have created new approaches for research into complex physical phenomena by use of numerical modeling tools. Despite these differences between modern day research and the situation faced by researchers in the eighteenth and nineteenth centuries, some properties of research activities are remarkably similar. A constant feature of scientific research is the need to conduct accurate experiments and develop more adequate model equations to describe the natural phenomena we observe. However, there is also a debate concerning the value of accuracy and practicality in describing these phenomena. While Airy and Stokes were debating the existence and basic properties of solitary waves of permanent shape in channels on theoretical grounds, Russell was constructing boats that were capable of high speed travel in channels, helped in part by this very wave phenomenon. To some extent, a similar debate is ongoing today within the tsunami research community, where on one side there is a need to develop models that represent fundamental properties of tsunami waves as accurately as possible in order to study the wave transformation and run-up processes in detail, and on the other side a need to develop tools for operational forecasting of tsunami wave events that are adequate and practical for warning purposes.


\section*{Acknowledgements}

This work was supported by ETAG grant PUT1378. Authors also thank the PHC PARROT project No 37456YM, which funded the authors’ visits to France and Estonia and allowed this collaboration.


\bibliography{Wave_mathematics_book}


\end{document}